\documentclass[aps,plb, nofootinbib, notitlepage,onecolumn,superscriptaddress]{revtex4-1}

\usepackage{bm}
\usepackage{latexsym}
\usepackage{amsmath,amsfonts,amssymb}
\usepackage{graphicx,epsfig}
\usepackage{psfrag}
\usepackage{amsthm}
\usepackage{color}
\usepackage{braket}
\usepackage{indent first}
\usepackage[normalem]{ulem}
\usepackage{hyperref}
\usepackage[titletoc,title]{appendix}
 \usepackage{environ}
\usepackage{mathtools}
\usepackage{float}
\usepackage[makeroom]{cancel}
\usepackage{bbold}
\usepackage{footmisc}
\usepackage{titlesec}

\usepackage[utf8]{inputenc}

\makeatletter
\newcommand\setcurrentname[1]{\def\@currentlabelname{#1}}
\makeatother

\begin{document}

\title{Test of Quantum Gravity in Optical Magnetometers}

\author{Mitja Fridman}
\email{fridmanm@uleth.ca}
\affiliation{Theoretical Physics Group and Quantum Alberta, Department of Physics and Astronomy,
University of Lethbridge,
4401 University Drive, Lethbridge,
Alberta, T1K 3M4, Canada} 

\author{James Maldaner}
\affiliation{Department of Physics, University of Alberta, 4-181 CCIS, Edmonton, Alberta, T6G 2E1, Canada} 

\author{Gil Porat}
\affiliation{Department of Physics, University of Alberta, 4-181 CCIS, Edmonton, Alberta, T6G 2E1, Canada}
\affiliation{Department of Electrical and Computer Engineering,
University of Alberta, 9211 116 St NW, Edmonton, Alberta T6G 1H9, Canada}

\author{Saurya Das}
\email{saurya.das@uleth.ca}
\affiliation{Theoretical Physics Group and Quantum Alberta, Department of Physics and Astronomy,
University of Lethbridge,
4401 University Drive, Lethbridge,
Alberta, T1K 3M4, Canada}

\begin{abstract}
In this work, quantum gravity effects, which can
potentially
be measured in magnetometers through the Larmor frequency of atoms in an external magnetic field, are estimated. 
It is shown that the thermal motion of atoms can, in principle, produce measurable quantum gravity effects, given the precision of modern magnetometers. If the particle velocities are caused by some other mechanism, such as convection, it is shown that the quantum gravity effects may be observed with the magnetometer's proposed detection threshold.
An actual state-of-the-art optical magnetometer experiment is being designed to search for these effects
and is described in a companion paper by Maldaner et al. (2023) \cite{magnetometer2023}.
\end{abstract}

\maketitle

\section{Introduction}


Although Quantum Theory (QT) and General Relativity (GR) are immensely successful theories in their 
own domains \cite{Zeilinger:1999zz,Will:2014kxa}, a truly successful and predictive theory of quantum gravity (QG) remains elusive. 
Several promising theories have been studied extensively,
such as String Theory \cite{ST1,ST2,ST3,ST4} and Loop Quantum Gravity \cite{Nicolai:2005mc,Thiemann:2006cf} among others, and while they differ significantly in their approaches and outcomes, they all seem to agree 
that a consistent theory of QG must fundamentally incorporate a minimum measurable length.
This modifies the Heisenberg Uncertainty Principle to the so-called Generalized Uncertainty Principle (GUP), whose effects can in turn be potentially tested in low-energy laboratory experiments \cite{Amelino-Camelia:2008aez}. 
This, on one hand, mitigates the need for building impractically large colliders to achieve the Planck energy scale 
($E_P \simeq 10^{16}\,\mathrm{TeV}$), where QG effects are generally believed to dominate, while on the other hand, offers a window for testing some of the robust predictions of all QG theories in laboratory based settings, such as the minimum length effects, by increasing the precision of measurements.


In this work, the linear and quadratic form of the GUP, consistent with all approaches to QG, and
first introduced in  
Refs. \cite{Ali:2009zq,Ali:2011fa}, 
as a modified $[x_i,p_j]$ commutator in $3$-dimensions
is considered  
%
 \begin{equation}
 \label{gup}
     [x_i,p_j]=i\,\hbar\left(\delta_{ij}-\alpha\left(p\,\delta_{ij}+\frac{p_i\,p_j}{p}\right)+\beta\left(p^2\,\delta_{ij}+3\,p_i\,p_j\right)\right)~, ~~i,j=1,2,3,
 \end{equation}
where $\alpha \equiv \alpha_0/(M_{P}\,c)$ and $\beta \equiv \beta_0/(M_{P}\,c)^2$, $M_p$ being the Planck mass 
($\simeq 2\times 10^{-8}$ kg), 
$\alpha_0$ and $\beta_0$ the dimensionless linear and quadratic GUP parameters, respectively, and $p=\sqrt{p_i\,p_i}$. Note that the GUP corrections are proportional to the Planck length, $\ell_P\simeq10^{-35}\,\mathrm{m}$, via $M_P\,c=\hbar/\ell_P$, where $\hbar$ is the reduced Planck constant, which shows that the GUP corrections are of QG origin.
One can see that the standard Quantum Mechanical (QM) operator for momentum $-i\,\hbar\,\partial_{i}$ cannot be used for $p_i$, because it does not satisfy the GUP commutation relation from Eq. (\ref{gup}). 
However, a set of canonical operators $x_{0i}$ and $p_{0i}$ can be defined, such that they satisfy the standard QM commutation relation $[x_{0i},p_{0j}]=i\,\hbar\,\delta_{ij}$.
Therefore, one can write
$p_{0i}=-i\,\hbar\,\partial_{x_{0i}}$. In terms of 
$x_{0i}$ and $p_{0i}$, one can define a transformation
 \begin{equation}
 \label{cops}
     x_i=x_{0i}\,\,\,\,\,\,\text{and}\,\,\,\,\,\,p_i=p_{0i}\,(1-\alpha\,p_0+2\,\beta\,p_0^2)~,
 \end{equation}
between physical and canonical operators, 
where $p_0=\sqrt{p_{0k}\,p_{0k}}$. 
%

This work predicts QG signatures in the Larmor frequency of atoms, parameterized by $\alpha_0$ and $\beta_0$. To search for these QG signatures,
a dedicated magnetometer experiment is being designed  \cite{magnetometer2023}. 
A magnetometer is a device which can measure magnetic fields or dipole moments of atoms. 
In particular, for the latter, the interactions of nuclear spins with external magnetic fields are measured. 
It turns out that \emph{optical magnetometers} are ideal for
testing fundamental physics, due to their high precision \cite{budker_jackson-kimball_2013}. In optical magnetometer experiments, one uses light to measure the response of angular momenta of atoms in an external magnetic field. 
In this context, the QG signatures are explored through GUP inspired modifications of the Larmor frequency of probed atoms. Specifically, QG modifications of the Larmor frequency of ${}^{129}\mathrm{Xe}$ atoms are considered in detail. The ${}^{129}\mathrm{Xe}$ species is chosen because it has non-zero nuclear spin, is stable and has a long relaxation time in normal conditions ($T=20^o\,\mathrm{C}$ and $p=1\,\mathrm{atm}$), which allows to make highly precise measurements. One of the methods to measure the Larmor frequency of such atoms is the two-photon laser spectroscopy, described in detail in Ref. \cite{Demtroder2002}.

Generic quantum gravity corrections to predictions of 
physical theories such as the standard model can be written as $\alpha_0\,\ell_P $ and $\sqrt{\beta_0}\, \ell_P$, which evidently defines two `new' length scales. 
 Since such a scale has not been observed at the LHC, which probes up to the electroweak scale, 
 given by $\ell_{EW}=\alpha_{EW}\ell_P$, where $\alpha_{EW}=10^{17}$,
  and it being reasonable to assume that the new scale is no smaller than the Planck scale, one gets the following inequality 
$1 \leq \alpha_{0},\sqrt{\beta_{0}} \leq 10^{17}$.
Now, there has been several studies which further tighten this bound. For example, considerations of quantum gravity effects in a scanning tunneling microscope lead to $\sqrt{\beta_0}<10^{10.5}$ \cite{Das:2008kaa}, while
that of gravity bar detectors give $\sqrt{\beta_0}<10^{16.5}$ \cite{Marin:2013}, interferometry experiments, $\sqrt{\beta_0}<10^{18}$ \cite{Luciano:2021cna}, and Bose-Einstein condensates, $\alpha_0, \sqrt{\beta_0}<10^{19}$ \cite{Das:2021skl}.
Note that the above bounds are all functions of the experimental precision, and that there are no predictions for $\alpha_0$ and $\beta_0$ as such. The only constraint is that obtained via the length scale comparisons above. 
This work, accompanied by Ref. \cite{magnetometer2023}, provides a window to directly probe the scale $\alpha_0,\sqrt{\beta_0}\sim 10^8$ using the above described magnetometer experiment.

The work is structured as follows. A brief summary of the design of the proposed experiment is given in Section \ref{sec:exp}, which is followed by a theoretical discussion on modifying the Larmor frequency by using GUP in Section \ref{sec:lfgup}.   QG signatures for thermally and non-thermally induced atom velocities are studied separately in Sections \ref{sec:thermalvel} and \ref{sec:nonthermalvel}, respectively. The work concludes with a summary of results in Section \ref{sec:conc}.

\section{The Experiment}
\label{sec:exp}

\begin{figure}[h]
\centering
\includegraphics[scale=0.8]{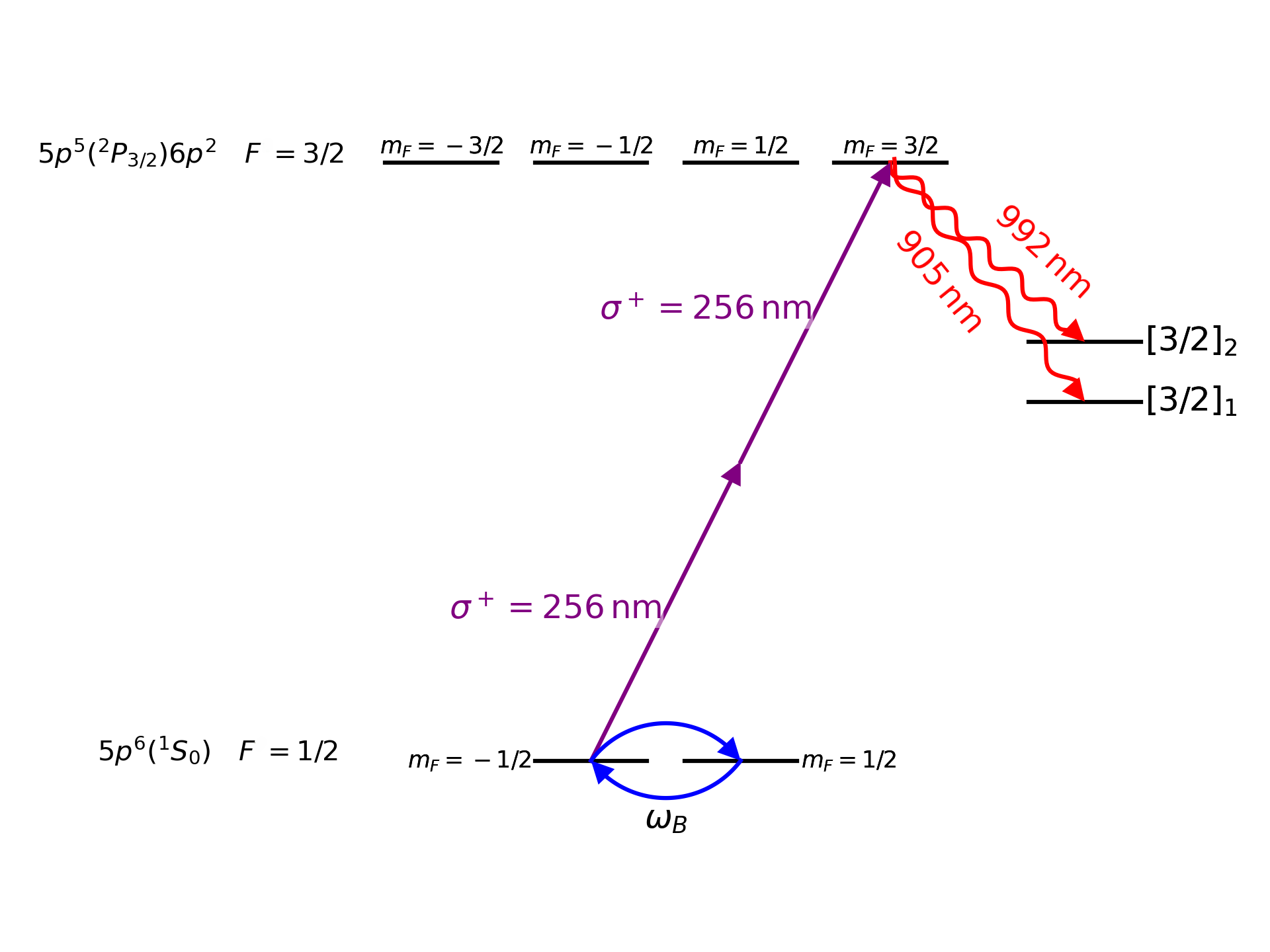}
\vspace{-0.8cm}
    \caption{Detailed energy level diagram of the proposed experiment. A two-photon absorption of circularly polarized DUV light excites the ${}^{129}\mathrm{Xe}$ atom to a state with $m_F=3/2$, which decays into one of the two intermediate states through IR emission. The external magnetic field induces Larmor precession with angular frequency $\omega_B$ between the hyperfine states with $m_F=-1/2$ and $m_F=1/2$.}
  \label{enlevs}
\end{figure}

The idea of the experiment is based on Refs. \cite{budker_jackson-kimball_2013,Demtroder2002} and described in detail in Ref. \cite{magnetometer2023}. The $^{129}\mathrm{Xe}$ atom in the ground state $5p^6\,{}^1\!S_0$,
with a total angular momentum quantum number $F=1/2$ ($F$ is defined through the operator sum $\mathbf{F}=\mathbf{J}+\mathbf{I}$, where $\mathbf{J}$ is the total electron angular momentum and $\mathbf{I}$ the total nuclear angular momentum) and projection $m_F=-1/2$, is excited by a circularly polarized deep ultra violet (DUV) light with $\sigma^+=256\,\mathrm{nm}$ at resonance to the state $5p^5\,({}^2\!P_{3/2})\,\,6p\,\,{}^{2}[5/2]_2$, which is a two-photon transition. This state then decays to one of two intermediate states $5p^5\,({}^2\!P_{3/2})\,\,6p\,\,{}^{2}[3/2]_{1,2}$, which emit IR photons with wavelengths $\lambda_{I\!R1}=905\,\mathrm{nm}$ and $\lambda_{I\!R2}=992\,\mathrm{nm}$, respectively, before decaying back to the ground state. The $^{129}\mathrm{Xe}$ atom has two ground state sublevels with $m_F=-1/2$ and $m_F=+1/2$, due to the hyperfine interaction. Since the $m_F=+1/2$ state cannot absorb the UV light, an ensemble of $^{129}\mathrm{Xe}$ atoms will eventually become spin polarized, where all atoms are in the same spin state of $m_F=+1/2$. Another possibility is to use the established method of spin-exchange optical pumping \cite{doi:10.1073/pnas.1306586110}. Then a uniform external magnetic field $\mathbf{B}$ is applied to induce the so-called \emph{Larmor precession} of $^{129}\mathrm{Xe}$ atoms. The atoms start to oscillate between the $m_F=+1/2$ and $m_F=-1/2$ states with the corresponding Larmor frequency. As the DUV light reinitiates the above excitation-decay process of atoms, which pass through the state $m_F=-1/2$, the IR emission oscillates with the exact Larmor frequency of the $^{129}\mathrm{Xe}$ atoms, which can be precisely measured. A detailed energy level diagram, highlighting the above experiment, is shown in Fig. \ref{enlevs}. A similar experiment, considering two-photon laser spectroscopy of an ensemble of ${}^{129}\mathrm{Xe}$ atoms, has been conducted by the authors in Ref. \cite{PhysRevA.97.012507}, where they use an atomic transition, different from the one described above.

To apply QG modifications to the Larmor frequency of an atom, one needs to consider contributions of electrons and the nucleus, which comprise the atom in question. From the ground state structure $5p^6\,{}^1\!S_0$ of the $^{129}\mathrm{Xe}$ atom, one can see that the total electron angular momentum $\mathbf{J}=0$, and the total angular momentum $\mathbf{F}=\mathbf{I}$. This means that the electrons do not induce their own magnetic dipole moment, and consequently do not contribute to the Larmor precession. Furthermore, the only cause of the magnetic dipole moment and the Larmor precession is the nuclear spin. To describe the interaction between the magnetic dipole moment of the nucleus and the external magnetic field $\mathbf{B}$, a suitably adapted Hamiltonian is considered \cite[p. 85, p. 152]{s2008introductory,landau}
\begin{eqnarray}
\label{hamiltoniannuc}
    H_N\!\!&=&\!\!H_{0N}-\mathbf{m}_N\cdot\mathbf{B}+\frac{e^2}{8\,m_p\,c^2}\sum_i(\mathbf{B\times\mathbf{r}_i})^2 \nonumber \\
    &\equiv&\!\!H_{0N}+H_{BN}+H_{1N}~,
\end{eqnarray}
where $H_{0N}$ is the Hamiltonian of the nucleus in the absence an external magnetic field, $\mathbf{m}_N$ the nuclear magnetic dipole moment operator, $e=+e_0$ the proton charge, $m_p$ the proton mass and $\mathbf{r}_i$ the radius of proton $i$ in the nucleus. The sum in the last term from Eq. (\ref{hamiltoniannuc}) goes over all protons $i$ in the nucleus. $H_{BN}$ and $H_{1N}$ in Eq. (\ref{hamiltoniannuc}) are defined for later convenience. It is useful to estimate the magnitudes of terms from Eq. (\ref{hamiltoniannuc}), to have an idea which effects are expected to contribute most in the following considerations. The expectation value of the leading order term $\langle H_{0N}\rangle$ is clearly the largest of the three \cite{s2008introductory}, and needs no further discussion in this context. However, the other two terms from Eq. (\ref{hamiltoniannuc}) need to be examined carefully. 

The magnitude of the $H_{BN}$ term can be estimated by taking an experimentally plausible external magnetic field $\mathbf{B}=2.7\,\mathrm{\mu T}\,\,\hat{z}$ \cite{magnetometer2023}, assumed parallel to the nuclear spin orientation, and by evaluating the expectation value of the nuclear magnetic dipole operator $\mathbf{m}_N$ in the ground state of $^{129}\mathrm{Xe}$. This expectation value turns out as $\langle \mathbf{m}_N\rangle_{Xe}=g_{Xe}\,\mu_{N}\,I\,\hat{z}$, where $g_{Xe}$ is the nuclear gyromagnetic factor for $^{129}\mathrm{Xe}$ and $\mu_{N}=e\,\hbar/2\,m_p$ the nuclear magneton \cite{s2008introductory}. The magnitude of $H_{BN}$ can then be estimated as
\begin{equation}
\label{hbn}
    |\langle H_{BN}\rangle|= |g_{Xe}\,\mu_{N}\,I\,B|=1.06\times10^{-32}\,\mathrm{J}~,
\end{equation}
where $g_{Xe}=-1.556~$ \cite{Makulski2015}, $I=F=1/2$ and $\mu_N=5.05\times10^{-27}\,\mathrm{J/T}$ were used. One can find the nuclear magnetic dipole moment of the $^{129}\mathrm{Xe}$ atom as $\mu_{Xe}=g_{Xe}\,\mu_{N}\,I=-0.778\,\mu_N=-3.93\times10^{-27}\,\mathrm{J/T}$. The magnitude of the $H_{1N}$ term can be estimated by the external magnetic field $\mathbf{B}$, assumed above, and a distribution of protons in the nucleus, which maximizes the contribution of this term. This provides a theoretical upper bound of the magnitude, which the $H_{1N}$ term can contribute. Such a distribution of protons, which maximizes the $H_{1N}$ term, is realized by assuming that all protons are located in a circle on the surface of the nucleus ($|\mathbf{r}_i|\sim R_{Xe}$ for all $i$, where $R_{Xe}$ is the radius of a $^{129}\mathrm{Xe}$ nucleus), where this circle is perpendicular to the external magnetic field, i.e., $\mathbf{r}_i\perp\mathbf{B}$ for all $i$, which maximizes the cross products from $H_{1N}$ in the ground state of $^{129}\mathrm{Xe}$. The radius of a stable $^{129}\mathrm{Xe}$ nucleus is approximated by $R_{Xe}=R_0\,A_{Xe}^{1\slash3}=6.06\times10^{-15}\,\mathrm{m}$, where $R_0=1.2\times10^{-15}\,\mathrm{m}$ is an experimentally determined nuclear scale, and $A_{Xe}=129$ the atomic mass number of $^{129}\mathrm{Xe}$ \cite{s2008introductory}. Since the sum in $H_{1N}$ goes only over the protons in the nucleus, there will be as many terms in the sum, as there is protons in the $^{129}\mathrm{Xe}$ nucleus, which is $Z_{Xe}=54$. The magnitude of $H_{1N}$ can then be estimated as
\begin{equation}
\label{h1n}
    |\langle H_{1N}\rangle|\lesssim\frac{e^2}{8\,m_p\,c^2}\,54\,B^2R_{Xe}^2=\frac{27\,e_0^2}{4\,m_p\,c^2}\,B^2R_{Xe}^2=3.08\times10^{-67}\,\mathrm{J}~,
\end{equation}
where the accepted values for $m_p$, $e_0$ and $c$ were used. One can then compare the contributions of the $H_{BN}$ and $H_{1N}$ terms, given by Eqs. (\ref{hbn}) and (\ref{h1n}), respectively, as
\begin{equation}
\label{hest}
    \frac{|\langle H_{1N}\rangle|}{|\langle H_{BN}\rangle|}\lesssim10^{-35}~,
\end{equation}
from where it can be seen that the contribution of the $H_{1N}$ term is more than 35 orders of magnitude smaller than the contribution of the $H_{BN}$ term. 
This implies the $H_{1N}$ term can be safely neglected in further considerations. Given the estimation of the GUP parameters obtained in the following, the $H_{1N}$ term is also negligible compared to the GUP correction terms.

\section{Larmor Frequency and GUP}
\label{sec:lfgup}

The QG corrections to the Larmor frequency of atoms,
motivated by GUP from Eq. (\ref{gup}), have been explored in detail in Ref. \cite{Bosso:2016frs}, where the analysis considers the Larmor precession of atoms, caused by the the total electron spin $\mathbf{J}$. However, for the purposes of the proposed ${}^{129}\mathrm{Xe}$ experiment, one needs to generalize this analysis to include nuclear spin, since $\mathbf{J}=0$ and $\mathbf{F}=\mathbf{I}$. It turns out that such a generalization is straightforward, following Ref. \cite{Bosso:2016frs}, where an arbitrary nucleus with a nuclear gyromagnetic factor $g_{nuc}$ and a nuclear magnetic dipole moment $\mu_{nuc}=g_{nuc}\,\mu_N\,I$ is considered. To achieve this, one considers the nuclear Hamiltonian from Eq. (\ref{hamiltoniannuc}), suitably modified by GUP through Eq. (\ref{cops}) and denoted by $H_{N}^{QG}$, while neglecting the $H_{1N}$ term. It turns out that the GUP corrected electron magnetic moment operator, as seen in Ref. \cite{Bosso:2016frs}, can be simply replaced by a GUP corrected nuclear magnetic moment operator since both obey the same modified spin algebra, as
\begin{eqnarray}
\mathbf{m}_N=\mathbf{m}_{N0}\,(1-\alpha\,p_0+\beta\,p_0^2)~,
\end{eqnarray}
where $\displaystyle{\mathbf{m}_{N0}=-\frac{g_{nuc}\,\mu_N}{\hbar}\,\mathbf{I}}$ and $p_0$ operators act on the nuclear wave function
\begin{eqnarray}
\Psi_N(\mathbf{r},t)=\psi_N(\mathbf{r},t)\,[a(t)\,|+\rangle_N+b(t)\,|-\rangle_N]~.
\end{eqnarray}
In the above, $\psi_N(\mathbf{r},t)$ is the spatial part and the linear combination $a(t)\,|+\rangle_N+b(t)\,|-\rangle_N$ is the spin part, with $a(t)$ and $b(t)$ time dependent functions, which satisfy $|a(t)|^2+|b(t)^2|=1$. For a quantization axis $\hat{z}$, parallel to $\mathbf{B}$, it turns out that $m_{Nz}\,|+\rangle_N=\mu_{nuc}\,|+\rangle_N$ and $m_{Nz}\,|-\rangle_N=-\mu_{nuc}\,|-\rangle_N$. Considering the above generalizations, one solves the Schrödinger equation of the nucleus $i\hbar\,\partial_t \Psi_N(\mathbf{r},t)=H_N^{QG}\,\Psi_N(\mathbf{r},t)$, which splits in two equations
\begin{eqnarray}
\label{sespa}
i\hbar\,\frac{\partial}{\partial\,t}\,\psi_N(\mathbf{r},t)=H_{0N}^{QG}\,\psi_N(\mathbf{r},t)~,
\end{eqnarray}
and
\begin{eqnarray}
\label{sespin}
i\hbar\,\frac{\partial}{\partial\,t}\,[a(t)\,|+\rangle_N+b(t)\,|-\rangle_N]=
-\mathbf{m}_{N0}\cdot\mathbf{B}\,(1-\alpha\,p_0+\beta\,p_0^2)\,[a(t)\,|+\rangle_N+b(t)\,|-\rangle_N]~.
\end{eqnarray}
It turns out that Eq. (\ref{sespa}) is not relevant in describing the Larmor precession, because it contains no information about the atom's spin, and is therefore ignored in the following considerations. However, it can provide QG corrections to the nuclear shell model, which can be a topic for a future project. On the other hand, the solutions of $a(t)$ and $b(t)$ of Eq. (\ref{sespin}) return harmonic solutions of the Larmor precession, which oscillate with a QG corrected Larmor frequency\footnote{For consistency with Ref. \cite{magnetometer2023}, the Larmor frequency is denoted by $\omega_B$ rather than $\omega_{L}$, as seen in Ref. \cite{Bosso:2016frs}. This notation is used in order to avoid confusion with the laser frequency $\omega_{l}$, introduced in Ref. \cite{magnetometer2023}.}
\begin{eqnarray}
\label{larmor}
\omega_B=-\frac{2\,\mu_{nuc}\,B}{\hbar}(1-\alpha\,\langle p\rangle+\beta\,\langle p^2\rangle)~,
\end{eqnarray}
where the expectation values of powers of momentum $p$, are related to the nuclear spin $\mathbf{I}$ of the atom. 
Note that the above Larmor frequency takes the same shape as in Ref. \cite{Bosso:2016frs}, with the difference that $\mu_{nuc}$ refers to the magnetic dipole moment of the nucleus, and $p$ is now related to the spin of the nucleus. Since $p$ represents a measure of momentum, related to $\mathbf{I}$, it can be interpreted as the linear momentum of the nucleus up to leading order (see Ref. \cite{Bosso:2018uus} for higher order corrections). Such a generalization can be done for any atom, where both, $\mathbf{J}\neq0$ and $\mathbf{I}\neq0$ cause the Larmor precession through the total spin $\mathbf{F}=\mathbf{J}+\mathbf{I}$. Eq. (\ref{larmor}) remains the same, where the total magnetic dipole moment is just the sum of the electron and nuclear magnetic dipole moments $\mu_0=\mu_{nuc}+\mu_{e}$, and $p$ can be interpreted in the same way.

The velocities of atoms in the proposed experiment are expected to be non-relativistic \cite{magnetometer2023}. Therefore, the expectation value of momentum and its square can be written in terms of the classical momentum in the laboratory frame as $\langle p\rangle=m\,\langle v\rangle$ and $\langle p^2\rangle=m^2\langle v^2\rangle$, respectively, where $m$ is the mass of the considered atom, $\langle v \rangle$ the expectation value of its velocity and $\langle v^2 \rangle$ the expectation value of the square of its velocity. Therefore, one can rewrite Eq. (\ref{larmor}) as
\begin{eqnarray}
\omega_B=\frac{2\,|\mu_0|\,B}{\hbar}(1-\alpha\,m\langle v\rangle+\beta\,m^2\langle v^2\rangle)~,
\label{larmorCorr}
\end{eqnarray}
where the absolute value $|\mu_0|$ is taken, without loss of generality, since only the magnitude of $\omega_B$ can be measured. For convenience, one can define the expectation value of the QG correction as $\langle C \rangle=\alpha\,m\langle v\rangle-\beta\,m^2\langle v^2\rangle$. The above Larmor frequency can then be simplified as
\begin{eqnarray}
\label{larmorc}
\omega_B={\gamma B}(1-\langle C\rangle)~,
\end{eqnarray}
where $\gamma=2|\mu_0|/\hbar$ is the gyromagnetic ratio. To discuss experimental implications of the above QG corrected Larmor frequency, one needs to quantify the magnitude of the QG effect $\langle C\rangle$ and compare it to the precision of the proposed experiment, quantified in the same manner. The magnitude of the relative QG correction is quantified as
\begin{eqnarray}
\label{accu1}
\left| \frac{\delta\omega_B}{\omega_{B,0}}\right|_{QG} = \langle C \rangle ~,
\end{eqnarray}
where $\omega_{B,0}={\gamma B}$. The above allows one to compare the QG corrections with the feasible precision of the proposed experiment \cite{magnetometer2023}
\begin{eqnarray}
\label{accu2}
\left|\frac{\delta\omega_B}{\omega_{B,0}}\right|_{exp} \simeq 10^{-15}~.
\end{eqnarray}
This sensitivity is achieved by increasing the total measurement time, given the available laser power (see Section IV. B. and Fig. 3 in Ref. \cite{magnetometer2023}). The idea behind the theoretical discussion is to optimize the parameters
$m$ and $v$, in order to bring the magnitude of the QG effect from Eq. (\ref{accu1}) as close as possible to the proposed experimental precision from Eq. (\ref{accu2}),
determine the constants $\alpha_0$, $\beta_0$ if such effects are measured and put strict bounds on them
if such effects are not measured. 
Such optimization provides an estimate of conditions at which QG effects are most likely to be detected. 
Since Eq. (\ref{larmorc}) is valid for any atom, different species of atoms with corresponding $m$ and $\mu_0$ are studied, while providing detailed estimates for the ${}^{129}\mathrm{Xe}$ atom. 

\section{Thermal Distribution of Atom Velocities}
\label{sec:thermalvel}

In this section, QG signatures in the Larmor frequency, using ensembles of atoms, are explored. Since the specific magnetometer experiment is proposed to be conducted in normal conditions \cite{magnetometer2023}, the thermal velocity distribution of an ensemble of chosen atoms with masses $m$, is the three dimensional Maxwell-Boltzmann (MB) distribution \cite{landau,pathria}
\begin{eqnarray}
\label{maxboldist}
f_v(v)=\left(\frac{m}{2\,\pi\,k_BT}\right)^{3\slash2}\!\exp{\left(-\frac{m\,v^2}{2\,k_BT}\right)}\,4\,\pi\,v^2~,
\end{eqnarray}
where $v \in [0,\infty)$, $k_B$ the Boltzmann constant and $T$ the temperature of the gas. In the following, QG signatures are discussed in terms of average thermal velocities of atoms, suggested by $\langle C\rangle$, and in terms of individual velocities of atoms, described by $\langle C\rangle\longrightarrow C$. Note that $\langle C\rangle$ corresponds to a single particle. Therefore, for a distribution, $C$ can take a range of values, as explained in detail later in this section.

\subsection{Average Thermal Velocities}

The considered version of QG corrections $\langle C\rangle=\alpha\,m\,\langle v\rangle-\beta\,m^2\langle v^2\rangle$ suggests that it is natural to take the average thermal velocity $\langle v\rangle$ and the mean square of the velocity $\langle v^2\rangle$ of an atom in a thermalized ensemble, obtained from the MB distribution, to estimate the magnitude of $\langle C\rangle$. For convenience, one can approximately parameterize the atom mass with the atomic mass number $A$, as $m=A\,m_p$ \cite{s2008introductory}. Using this parameterization, one can write the average thermal velocity of species $A$ at temperature $T$ as
\begin{figure}[b]
\hbox{\hspace{-0.5cm}\includegraphics[scale=0.85,trim={1.5cm 2.8cm 0.0cm 3.8cm},clip]{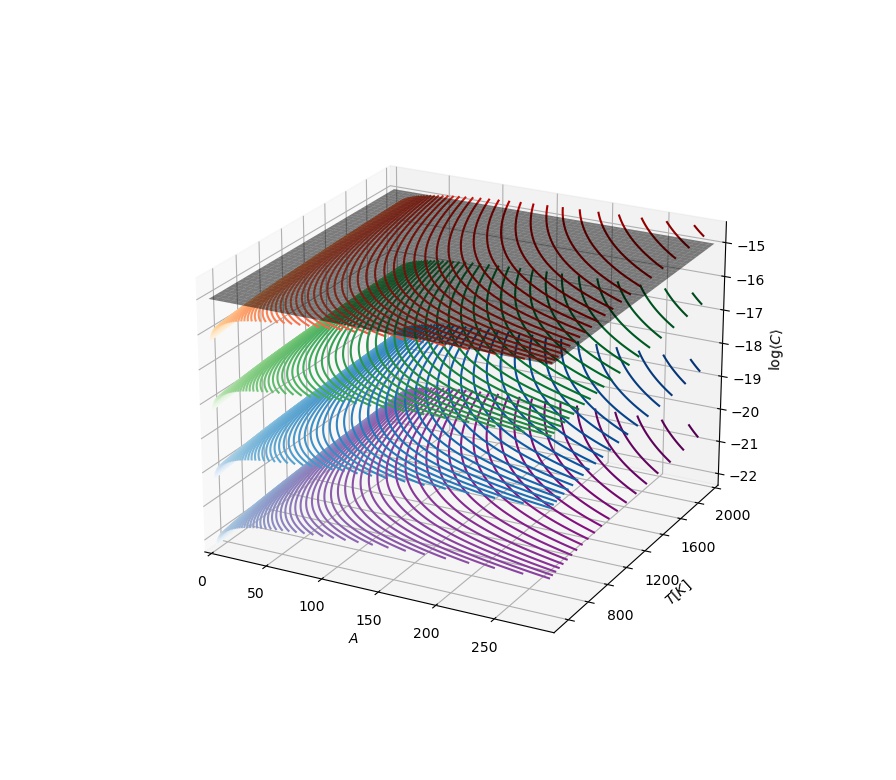}}
    \caption[Logarithmic dependence of $\langle C\rangle$ on mass number $A$ and temperature $T$ for four different values of $\alpha_0$]{Logarithmic dependence of $\langle C\rangle$ on mass number $A$ and temperature $T$ for four different values of $\alpha_0$; purple: $\alpha_0=10^2$, blue: $\alpha_0=10^4$, green: $\alpha_0=10^6$ and red: $\alpha_0=10^8$. The black flat surface is the experimental precision limit.}
    \label{logC11}
\end{figure}
\begin{eqnarray}
\label{maxv}
\langle v\rangle=\int_0^\infty v\,f_v(v)\,\mathrm{d}v \!\!&=&\!\!\sqrt{\frac{8\,k_B T}{\pi\, m}} \nonumber \\
&=&\!\!\frac{1}{\sqrt{A}}\,\sqrt{\frac{8\,k_BT_{RT}}{\pi\,m_p}}\,\sqrt{\frac{T}{T_{RT}}} \nonumber \\
&=&\!\!\frac{1}{\sqrt{A}}\,\langle v_H\rangle_{RT} \sqrt{\frac{T}{T_{RT}}}~,
\end{eqnarray}
and the mean square of the velocity as
\begin{eqnarray}
\label{maxv2}
\langle v^2\rangle=\int_0^\infty v^2f_v(v)\,\mathrm{d}v\!\!&=&\!\!\frac{3\,k_BT}{m} \nonumber \\
&=&\!\!\frac{1}{A}\,\frac{3\,k_BT_{RT}}{m_p}\,\frac{T}{T_{RT}} \nonumber \\
&=&\!\!\frac{1}{A}\,\langle v_H^2\rangle_{RT}\,\frac{T}{T_{RT}}~,
\end{eqnarray}
which are given in terms of the average velocity and the mean square of the velocity of an ensemble of Hydrogen atoms at room temperature $T_{RT}=293\,\mathrm{K}$, namely 
$\langle v_H\rangle_{RT} \approx 2480\,\,\mathrm{m/s}$ and $\langle v_H^2\rangle_{RT} \approx 7.256\times10^{6}\,\mathrm{m^2/s^2}$, respectively. Plugging Eqs. (\ref{maxv}) and (\ref{maxv2}) in the definition of $\langle C\rangle$, and using the above parameterization of $m$ in terms of $A$, one obtains 
\begin{eqnarray}
\label{maxcorr}
\langle C \rangle(\alpha_0, \beta_0;\,A, T)= \alpha_0\,\frac{\sqrt{A}\,m_p\,\langle v_{H}\rangle_{RT}}{M_{P}\,c}\,\sqrt{\frac{T}{T_{RT}}}-\beta_0\,\frac{A\,m_p^2\langle v_H^2\rangle_{RT}}{(M_P\,c)^2}\,\frac{T}{T_{RT}}~,
\end{eqnarray}
where the definitions of $\alpha$ and $\beta$, found below Eq. (\ref{gup}), have been used. For clarity, the logarithm of Eq. (\ref{maxcorr}) is shown in Fig. \ref{logC11}, as a monotonically increasing function of both, $A$ and $T$, where $\beta_0=\alpha_0^2$ is assumed. The dependence is shown for several different values of $\alpha_0$. The black flat surface corresponds to the experimental precision. One can see that QG signatures can be detected if $\alpha_0\approx10^8$. In this case, the detection of QG signatures, using ${}^{129}\mathrm{Xe}$ atoms with $A=A_{Xe}=129$, would take place at $T\approx560\,\mathrm{K}$, given Eq. (\ref{maxcorr}).

\subsection{Distribution of Thermal Velocities}

In the above, it was assumed that the expectation values in $\langle C\rangle$ refer to the averages of the whole distribution of atoms. However, Eq. (\ref{larmorc}) is derived for a single atom, which suggests that the expectation values $\langle p\rangle$ and $\langle p^2\rangle$ correspond to the single particle momentum $p$ and its square $p^2$ for that atom, and not to the averages of the whole ensemble. This quantum-classical correspondence then implies $\langle C \rangle \longrightarrow C$, $\langle v \rangle \longrightarrow v$ and $\langle v^2 \rangle \longrightarrow v^2$. Therefore, it turns out that atoms are distributed over a range of Larmor frequencies $\omega_B$ (over a range of $C$ as well), since they are MB distributed over all velocities $v\in[0,\infty)$, following Eq. (\ref{maxboldist}), resulting in a QG induced broadening of the Larmor frequency distribution. To see this, one rewrites Eq. (\ref{larmorCorr}), considering the single atom correspondence and atomic mass number parameterization, as
\begin{eqnarray}
\label{larmor-vel}
\omega_B=\omega_{B,0}\left(1-\alpha\,A\,m_p\,v+\beta\,A^2m_p^2v^2\right)~.
\end{eqnarray}
For convenience, one can also rewrite the MB distribution from Eq. (\ref{maxboldist}) in terms of the atomic mass number $A$ as 
\begin{eqnarray}
\label{mbd2}
f_v(v)=\left(\frac{m_p}{2\,\pi\,k_BT}\right)^{3\slash2}\!\!A^{3\slash2}\,\exp{\left(-\frac{A\,m_p\,v^2}{2\,k_BT}\right)}\,4\,\pi\,v^2~.
\end{eqnarray}
To obtain a QG corrected distribution of atoms over Larmor frequencies $f_\omega(\omega_B)$, given Eqs. (\ref{larmor-vel}) and (\ref{mbd2}), one makes the following change of variables
\begin{eqnarray}
\label{diff2}
f_v(v)\,\mathrm{d}v=f_\omega(\omega_B)\,\mathrm{d}\omega_B\,\,\,\,\implies\,\,\,\, f_\omega(\omega_B)=f_v(v)\left|\frac{\mathrm{d}v}{\mathrm{d}\omega_B}\right|~.
\end{eqnarray}
The derivative $|{\mathrm{d}v}/{\mathrm{d}\omega_B}|$ on the right-hand side of Eq. (\ref{diff2}) is obtained by solving Eq. (\ref{larmor-vel}) for
\begin{eqnarray}
\label{larvelcorr}
v=\frac{\alpha}{2\,\beta\,A\,m_p}\,\left(1-\sqrt{1+4\,\frac{\beta}{\alpha^2}\left(\frac{\omega_B}{\omega_{B,0}}-1\right)}\right)~,
\end{eqnarray}
where only the $-$ solution was considered, since the $+$ solution does not recover the standard result $\omega_B=\omega_{B,0}$ for $v=0$, and deriving $v$ over $\omega_B$, which then reads as
\begin{eqnarray}
\label{larvelder}
\left|\frac{\mathrm{d}v}{\mathrm{d}\omega_B}\right|=\frac{1}{\alpha\,A\,m_p\,\omega_{B,0}}\frac{1}{\sqrt{1+4\,\frac{\beta}{\alpha^2}\,\left(\frac{\omega_B}{\omega_{B,0}}-1\right)}}~.
\end{eqnarray}
By using Eqs. (\ref{larvelcorr}) and (\ref{larvelder}) in the change of variables from Eq. (\ref{diff2}), the distribution of atoms over $\omega_B$, is obtained as

\begin{eqnarray}
\label{omdist}
f_\omega(\omega_B)\!\!&=&\!\!\left(\frac{1}{2\,\pi\,A\,m_p\,k_BT}\right)^{3/2}\!\exp{\left(-\frac{\alpha_0^2(M_P\,c)^2}{\beta_0^2}\,\frac{\left(1-\sqrt{1+4\,\frac{\beta_0}{\alpha_0^2}\left(\frac{\omega_B}{\omega_{B,0}}-1\right)}\right)^2}{8\,A\,m_p\,k_BT}\right)} \nonumber \\
&\times&\!\!\pi\,\frac{\alpha_0\,(M_P\,c)^3}{\beta_0^2\,\omega_{B,0}}\,\frac{\left(1-\sqrt{1+4\,\frac{\beta_0}{\alpha_0^2}\left(\frac{\omega_B}{\omega_{B,0}}-1\right)}\right)^2}{\sqrt{1+4\,\frac{\beta_0}{\alpha_0^2}\left(\frac{\omega_B}{\omega_{B,0}}-1\right)}}~,
\end{eqnarray}
where the definitions of $\alpha$ and $\beta$, found below Eq. (\ref{gup}), have been used. From the above distribution, one can obtain the standard phenomenological quantities, similarly as seen with the MB distribution, such as the most probable Larmor frequency (at the peak of the distribution; $\mathrm{d}f_\omega/\mathrm{d}\omega_B=0$)
\begin{eqnarray}
\label{ompeak}
\omega_{B,\mathrm{peak}}=\omega_{B,0}\left(1-\alpha_0\,\frac{\sqrt{2\,A\,m_p\,k_BT}}{M_P\,c}+\beta_0\,\frac{2\,A\,m_p\,k_BT}{(M_P\,c)^2}\right)~,
\end{eqnarray}
the average Larmor frequency
\begin{eqnarray}
\label{omcorr}
\langle\omega_B\rangle=
    \int_{\omega_{B,0}}^{\infty}\!\omega_B\,\,f_\omega(\omega_B)\,\mathrm{d}\omega_B=\omega_{B,0}\left(1-\alpha_0\,\frac{\sqrt{{8}\,A\,m_p\,k_BT}}{\sqrt{{\pi}}\,M_P\,c}+\beta_0\,\frac{3\,A\,m_p\,k_BT}{(M_P\,c)^2}\right)~,
\end{eqnarray}
and the mean square of the Larmor frequency
\begin{eqnarray}
\langle\omega_B^2\rangle\!\!&=&\!\!\int_{\omega_{B,0}}^\infty\!\omega_B^2\,\,f_\omega(\omega_B)\,\mathrm{d}\omega_B \nonumber \\
&=&\!\!\omega_{B,0}^2\left(1-\alpha_0\,\frac{2}{M_P\,c}\sqrt{\frac{8\,A\,m_p\,k_BT}{\pi}}+(\alpha_0^2+2\,\beta_0)\frac{3\,A\,m_p\,k_BT}{(M_P\,c)^2}\right. \nonumber \\
&{}&\,\,\,\,\,\,\,\,\,\,\,\,\,\,\,\,\left.-\alpha_0\,\beta_0\,\sqrt{\frac{8}{\pi}}\,\frac{8\,(A\,m_p\,k_BT)^{3/2}}{(M_P\,c)^3}+\beta_0^2\,\frac{15\,A^2\,m_p^2\,k_B^2T^2}{(M_P\,c)^4}\right)~.
\end{eqnarray}
One can see that the correction terms in Eq. (\ref{omcorr}) correspond exactly to Eq. (\ref{maxcorr}), if expressed in terms of $\langle v_H\rangle_{RT}$ and $\langle v_H^2\rangle_{RT}$. However, this consideration provides more information, since now one has the distribution $f_\omega(\omega_B)$.

For detailed phenomenological research, where different GUP models are considered, one can straightforwardly reduce Eq. (\ref{omdist}) to obtain $f_\omega(\omega_B)$ for such models. For the linear GUP model ($\beta_0\longrightarrow0$), the above distribution turns out as
\begin{eqnarray}
\label{lindist}
    f_\omega^{\mathrm{Lin}}(\omega_B)\!\!&=&\!\!\left(\frac{1}{2\,\pi\,A\,m_p\,k_BT}\right)^{3\slash2}\!\exp{\left(-\frac{(M_P\,c)^2}{\alpha_0^2}\,\frac{\left(\frac{\omega_B}{\omega_{B,0}}-1\right)^2}{2\,A\,m_p\,k_BT}\right)} \nonumber \\
    &\times&\!\!4\,\pi\,\frac{(M_P\,c)^3}{\alpha_0^3\,\omega_{B,0}}\,\left(\frac{\omega_B}{\omega_{B,0}}-1\right)^2~, \label{lindist} 
\end{eqnarray}
for which $\omega_{B,\mathrm{peak}}^{\mathrm{Lin}}$ and $\langle\omega_B\rangle^{\mathrm{Lin}}$ are those from Eqs. (\ref{ompeak}) and (\ref{omcorr}), respectively, when $\beta_0\longrightarrow0$. 
Next, for the quadratic GUP model ($\alpha_0\longrightarrow0$), the distribution from Eq. (\ref{omdist}) turns out as 
\begin{eqnarray}
\label{quaddist}
f_\omega^{\mathrm{Quad}}(\omega_B) 
\!\!&=&\!\!\left(\frac{1}{2\,\pi\,A\,m_p\,k_BT}\right)^{3\slash2}\!\exp{\left(-\frac{(M_P\,c)^2}{\beta_0}\,\frac{\left(\frac{\omega_B}{\omega_{B,0}}-1\right)}{2\,A\,m_p\,k_BT}\right)} \nonumber \\
&\times&\!\!2\,\pi\,\frac{(M_P\,c)^3}{\beta_0^{3/2}\,\omega_{B,0}}\,\sqrt{\frac{\omega_B}{\omega_{B,0}}-1}~, \label{quaddist}
\end{eqnarray}
for which $\omega_{B,\mathrm{peak}}^{\mathrm{Quad}}$ and $\langle\omega_B\rangle^{\mathrm{Quad}}$ are those from Eqs. (\ref{ompeak}) and (\ref{omcorr}), respectively, when $\alpha_0\longrightarrow0$.

From Eqs. (\ref{ompeak}) and (\ref{omcorr}) one can see that the peak and mean of the Larmor frequencies 
are shifted with respect to the Larmor frequency $\omega_{B,0}$ in standard theory, as long as $T>0$.
Furthermore, Eq. (\ref{omdist}) implies a broadening of the measured Larmor frequency distribution, due to thermal motion of atoms. With increasing $\alpha_0$ and $\beta_0$, the deviation of $\omega_B$ from $\omega_{B,0}$ gets larger and the width of the distribution gets broader, and vice-versa. Increasing the temperature $T$ and the atomic mass number $A$, provide the same effect on the distribution. Since the distribution of QG corrections, described by Eq. (\ref{omdist}), is so close to $\omega_{B,0}$, the sampling of such small steps of $\omega_B$ is not possible with current computing power. Therefore, it is more convenient to visualize the distribution of atoms over QG corrections $C$
instead, where
\begin{eqnarray}
\label{c-vel}
C=\alpha_0\,\frac{A\,m_p\,v}{M_P\,c}-\beta_0\,\frac{A^2m_p^2v^2}{(M_P\,c)^2}=\left|\frac{\delta\omega_B}{\omega_{B,0}}\right|\equiv\left|\frac{\delta\omega}{\omega}\right|
\end{eqnarray}
is the new distribution variable. To achieve this, the same procedure used to derive Eq. (\ref{omdist}) is followed. The corresponding change of variables is
\begin{eqnarray}
f_C(C)=f_v(v)\left|\frac{\mathrm{d}v}{\mathrm{d}C}\right|~,
\end{eqnarray}
which implies the replacement
\begin{eqnarray}
\frac{\omega_B}{\omega_{B,0}}-1\implies\frac{\delta\omega}{\omega}
\end{eqnarray}
of all such factors in distributions from Eqs. (\ref{omdist}), (\ref{lindist}) and (\ref{quaddist}), and does not include the $\omega_{B,0}$ factor in the denominator. These distributions then respectively read as 
\begin{eqnarray}
\label{Cdistr}
f_C\left(\frac{\delta\omega}{\omega}\right)\!\!&=&\!\!\left(\frac{1}{2\,\pi\,A\,m_p\,k_BT}\right)^{3/2}\!\exp{\left(-\frac{\alpha_0^2(M_P\,c)^2}{\beta_0^2}\,\frac{\left(1-\sqrt{1+4\,\frac{\beta_0}{\alpha_0^2}\left(\frac{\delta\omega}{\omega}\right)}\right)^2}{8\,A\,m_p\,k_BT}\right)} \nonumber \\
&\times&\!\!\pi\,\frac{\alpha_0\,(M_P\,c)^3}{\beta_0^2}\,\frac{\left(1-\sqrt{1+4\,\frac{\beta_0}{\alpha_0^2}\left(\frac{\delta\omega}{\omega}\right)}\right)^2}{\sqrt{1+4\,\frac{\beta_0}{\alpha_0^2}\left(\frac{\delta\omega}{\omega}\right)}}~,
\end{eqnarray}
for $\alpha_0,\beta_0\neq0$,
\begin{figure}[b]
    \hbox{\hspace{3.2cm}\includegraphics[scale=0.1815]{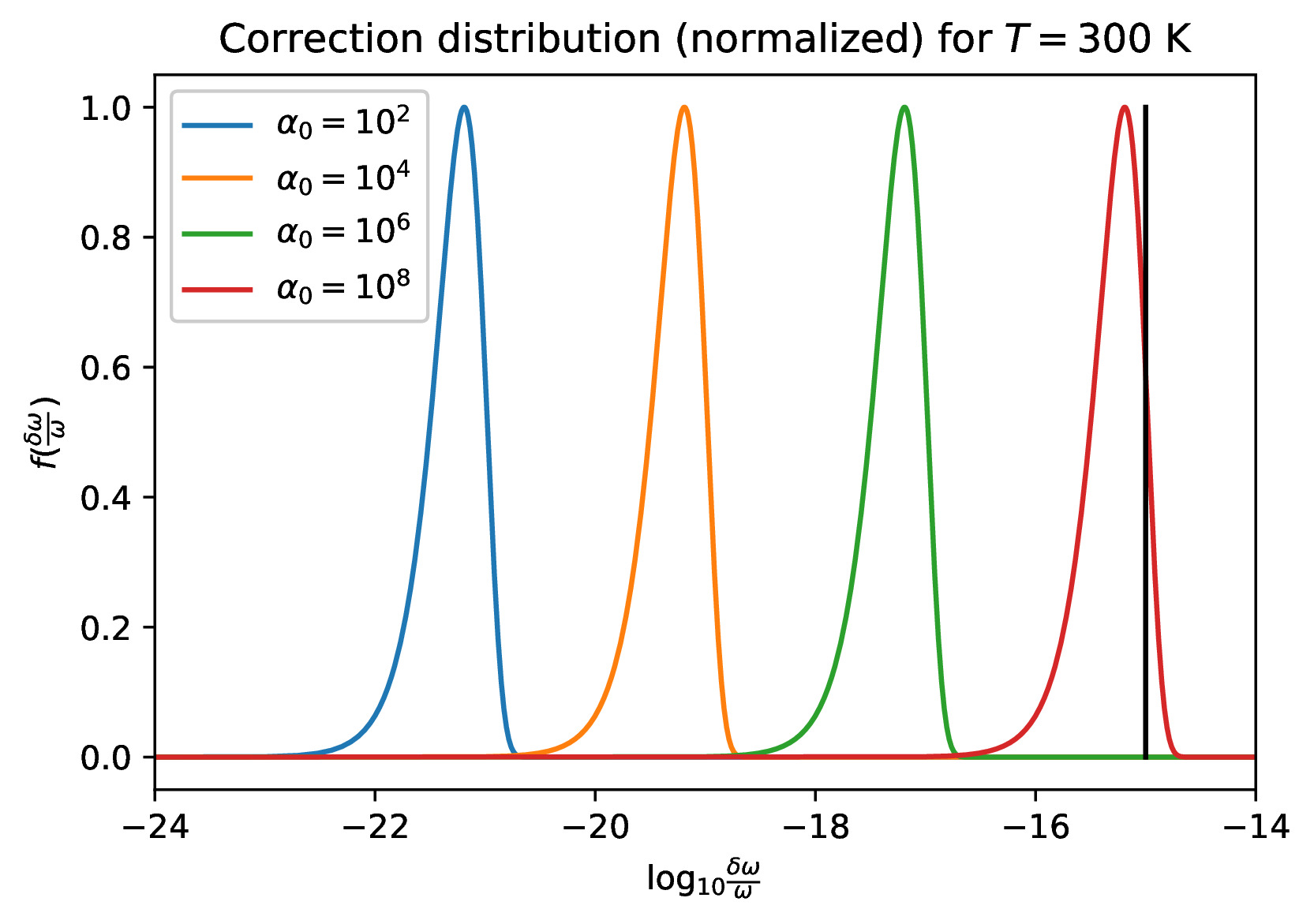}}
    \vspace{0.8cm}
    
    \hbox{\hspace{3.3cm}\includegraphics[scale=0.170]{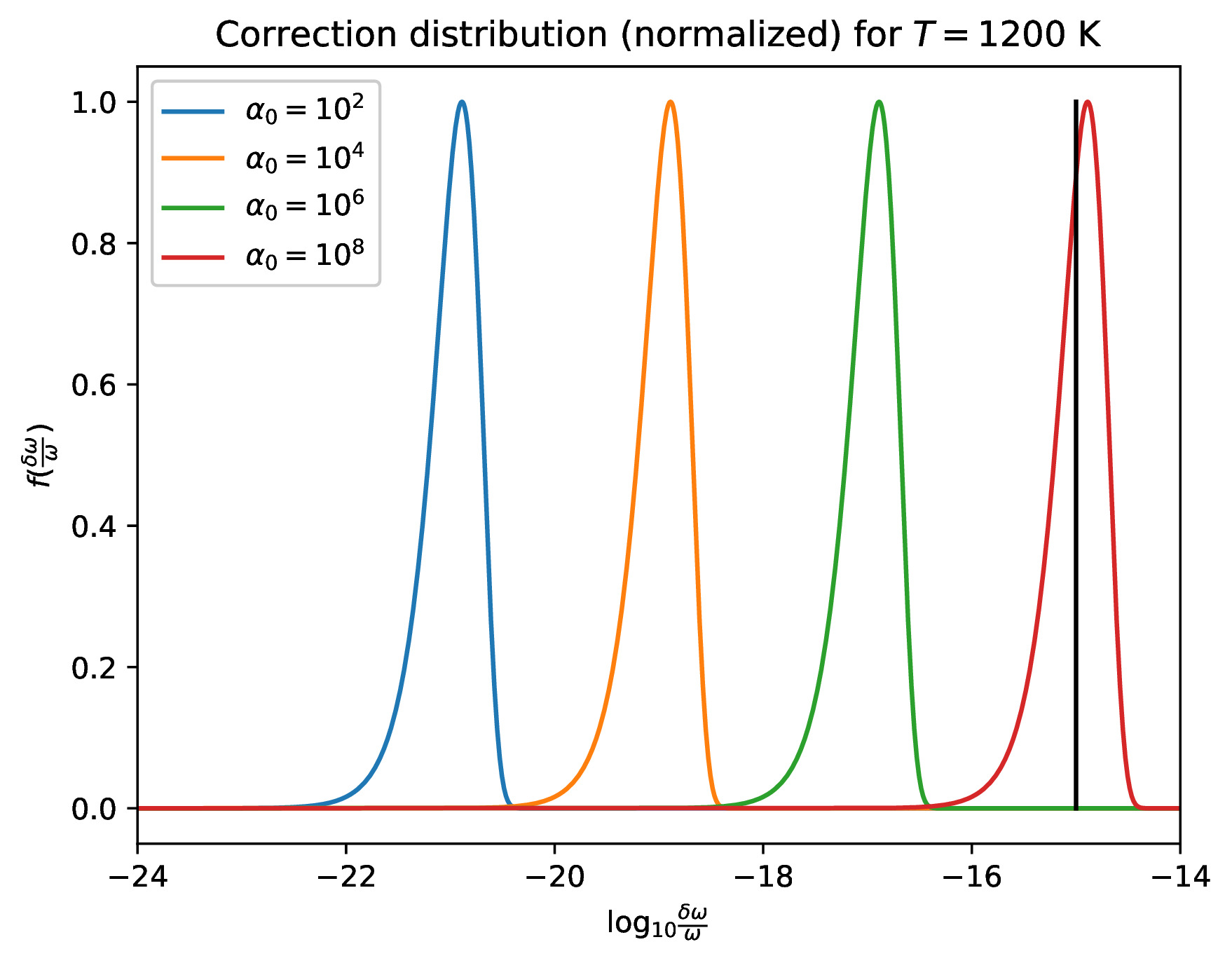}}
    \vspace{0.0cm}
    \caption[Distributions of ${}^{129}\mathrm{Xe}$ atoms over $\delta\omega\slash\omega$ for different values of $\alpha_0$]{Distributions of ${}^{129}\mathrm{Xe}$ atoms over $\delta\omega\slash\omega$ for different values of $\alpha_0$, at temperatures $T=300\,\mathrm{K}$ (top) and $T=1200\,\mathrm{K}$ (bottom). The black vertical line is the experimental precision limit.}
  \label{cdist1}
\end{figure}
\begin{eqnarray}
\label{linbroad}
    f_C^{\mathrm{Lin}}\left(\frac{\delta\omega}{\omega}\right)\!\!&=&\!\!\left(\frac{1}{2\,\pi\,A\,m_p\,k_BT}\right)^{3\slash2}\!\exp{\left(-\frac{(M_P\,c)^2}{\alpha_0^2}\,\frac{\left(\frac{\delta\omega}{\omega}\right)^2}{2\,A\,m_p\,k_BT}\right)} \nonumber \\
    &\times&\!\!4\,\pi\,\frac{(M_P\,c)^3}{\alpha_0^3}\,\left(\frac{\delta\omega}{\omega}\right)^2~,
\end{eqnarray}
for $\alpha\neq0$ and $\beta_0=0$, and
\begin{eqnarray}
\label{quadbroad}
f_C^{\mathrm{Quad}}\left(\frac{\delta\omega}{\omega}\right)\!\!&=&\!\!\left(\frac{1}{2\,\pi\,A\,m_p\,k_BT}\right)^{3\slash2}\!\exp{\left(-\frac{(M_P\,c)^2}{\beta_0}\,\frac{\frac{\delta\omega}{\omega}}{2\,A\,m_p\,k_BT}\right)} \nonumber \\
&\times&\!\!2\,\pi\,\frac{(M_P\,c)^3}{\beta_0^{3/2}}\,\sqrt{\frac{\delta\omega}{\omega}}~,
\end{eqnarray}
for $\alpha_0=0$ and $\beta_0\neq0$. The QG induced broadening, implied by Eq. (\ref{Cdistr}), for an ensemble of ${}^{129}\mathrm{Xe}$ atoms, $A=129$, is shown in Fig. \ref{cdist1}, for different values of $\alpha_0$ and temperatures $T=300\,\mathrm{K}$ (top) and $T=1200\,\mathrm{K}$ (bottom), where $\beta_0=\alpha_0^2$ was assumed. For various choices of the parameter $\alpha_0$, the heights of the peaks of the distributions differ by several orders of magnitude. Therefore, to 
present and compare them in the same figure, they are normalized, such that each peak assumes a
maximum value of unity. Notice that for $\alpha_0 \gtrsim  10^8$, the peak is close to, or even crosses the 
projected sensitivity of the magnetometer. Fig. \ref{cdist1} can be interpreted in the following way. The horizontal axis represents the magnitude of QG corrections, while the vertical axis represents the fraction of atoms. Given a thermal distribution of atoms and $\alpha_0$, the average magnitude of the QG signature will be localized around the peak of the distribution, since most of the atoms display this deviation, due to their movement with speeds near the average speed $\langle v\rangle$. Furthermore, since the horizontal axis is in a logarithmic scale, the broadening of the distribution is much higher for larger values of $\alpha_0$ and $T$.

For the proposed experimental sensitivity, the increase in temperature to $T=1200\,\mathrm{K}$ will increase the Doppler width by a factor of 2 (see Eq. (7) in Ref. \cite{magnetometer2023}) as compared to the $T=300\,\mathrm{K}$ case since it scales as the square root of the temperature. The number density of the Xenon gas will be scaled by a factor of $1/4$ (see paragraph before Eq. (9) in Ref. \cite{magnetometer2023}) if the same pressure is maintained. The optimal beam waist is scaled by a factor of $1/2^{1/4}$ (see Eqs. (C2), (C3) and (C6) in Ref. \cite{magnetometer2023}), so the beam area is reduced by a factor of $1/\sqrt{2}$. The number of Xenon atoms involved in the interaction is scaled by $1/4\sqrt{2}$ (see sentence after Eq. (13) in Ref. \cite{magnetometer2023}). This scales the average detection rate by a factor of $1/4\sqrt{2}$ (see Eq. (9) in Ref. \cite{magnetometer2023}). The effects of the Doppler width and the beam waist adjustments are canceled out for the two-photon transition rate, so it is not affected (see Eq. (8) in Ref. \cite{magnetometer2023}). The temperature increase also increases the shot noise variance by $4\sqrt{2}$  (see Eq. (12) in Ref. \cite{magnetometer2023}), as well as the projection noise variance by a factor of $4\sqrt{2}$ (see Eq. (13) in Ref. \cite{magnetometer2023}). Therefore, the fractional sensitivity is degraded by a factor of $2^{7/4}$  (see Eq. (11) in Ref. \cite{magnetometer2023}).

\begin{figure}[H]
\hbox{\hspace{2.8cm}\includegraphics[scale=0.1815]{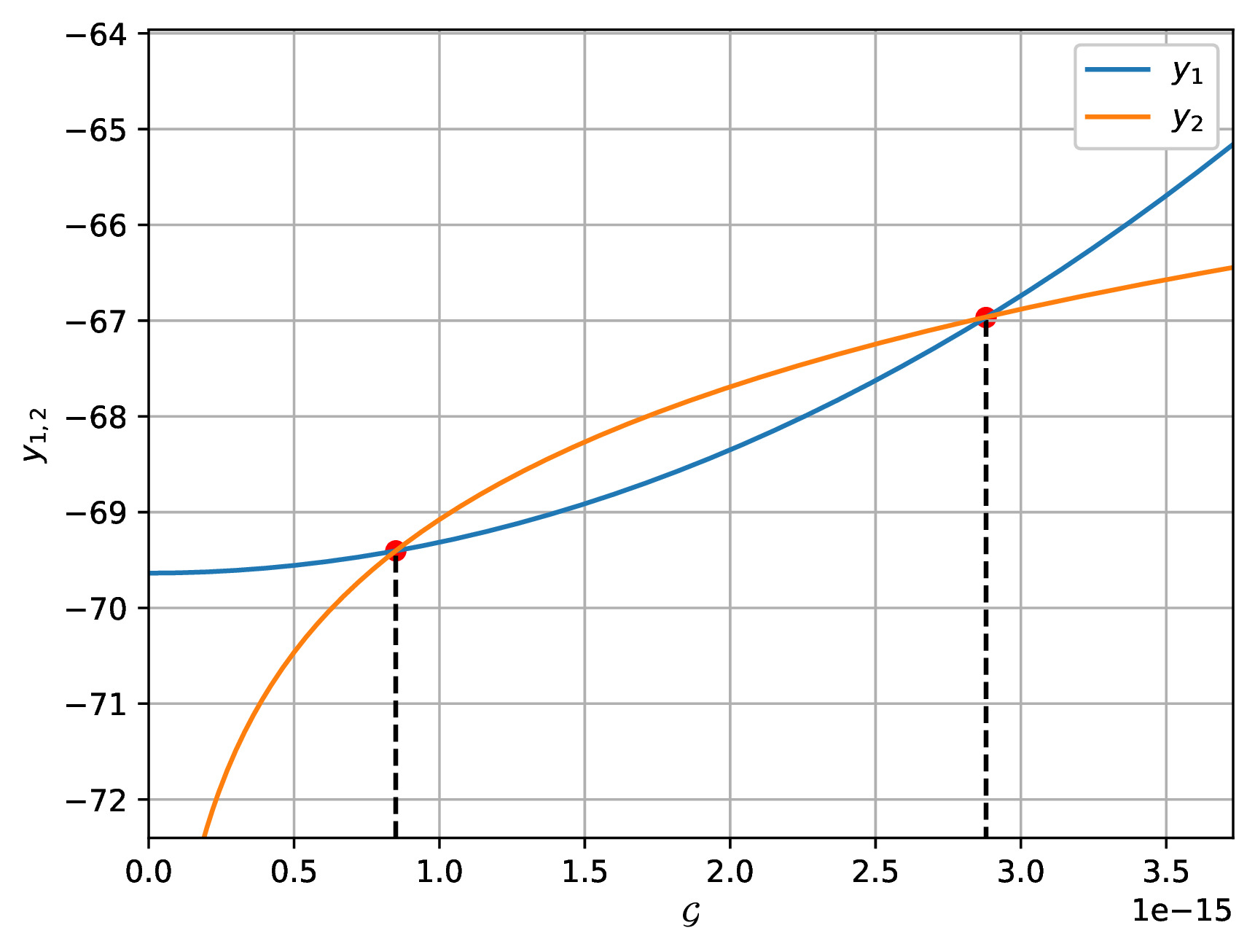}}
    \caption[Intersection points of $y_1$ and $y_2$]{Intersection points of $y_1$ and $y_2$, for an example of an ensemble of ${}^{129}\mathrm{Xe}$ atoms. $A=129$, at $T=560\,\mathrm{K}$ for $\alpha_0=10^8$.}
    \label{csolut}
\end{figure}

For a better visualization of the broadening, the measures of the second moment (standard deviation) and the Full Width Half Maximum (FWHM) of the distribution are considered. The second moment of the distribution from Eq. (\ref{omdist}) is 
\begin{eqnarray}
\label{smlin}
\sigma_{\omega}^2\!\!&=&\!\!\int_{\omega_{B,0}}^{\infty}(\omega_B-\langle\omega_B\rangle)^2f_\omega(\omega_B)\,\mathrm{d}\omega_B \nonumber \\
&=&\!\!\alpha_0^2\,\left(3-\frac{8}{\pi}\right)\frac{\omega_{B,0}^2\,A\,m_p\,k_BT}{(M_P\,c)^2}-\alpha_0\,\beta_0\,\sqrt{\frac{8}{\pi}}\,\frac{2\,\omega_{B,0}^2(A\,m_p\,k_BT)^{3/2}}{(M_P\,c)^3} \nonumber \\
&+&\!\!\beta_0^2\,\frac{6\,\omega_{B,0}^2\,A^2m_p^2k_B^2T^2}{(M_P\,c)^4}~.
\end{eqnarray}
From the above one can see that such broadening is caused entirely by QG effects, since as $\alpha_0,\beta_0\longrightarrow0$, the $\sigma_\omega^2$ vanishes. On the other hand, the FWHM of the distribution is obtained as 
\begin{eqnarray}
\label{fwhmgupmag}
\mathrm{FWHM}_\omega=\omega_{B_{2}}^{\mathrm{FWHM}}-\omega_{B_{1}}^{\mathrm{FWHM}}~,
\end{eqnarray}
where $\omega_{B_{1}}^{\mathrm{FWHM}}$ and $\omega_{B_{2}}^{\mathrm{FWHM}}$ are values, corresponding to both sides of the distribution at the half of its maximum. 
These values are obtained from Eq. (\ref{omdist}), by solving the following equation 
\begin{eqnarray}
\label{fwhm}
    f_\omega\left(\omega_{B_{1,2}}^{\mathrm{FWHM}}\right)=\frac{f_\omega\left(\omega_{B,\mathrm{peak}}\right)}{2}~.
\end{eqnarray}
The above has no closed form solution for $\omega_{B_{1}}^{\mathrm{FWHM}}$ and $\omega_{B_{2}}^{\mathrm{FWHM}}$. However, it is still possible to obtain these values numerically, by the following procedure. A short manipulation of Eq. (\ref{fwhm}) provides a transcendental equation 

\begin{eqnarray}
\label{transcendental}
&{}&\!\!\!\!\!\!\!\!\!\!\frac{\alpha_0^2\,(M_P\,c)^2}{\beta_0^2}\,\frac{\mathcal{G}_{1,2}^2}{8\,A\,m_p\,k_BT} +\ln{\left(4\,\frac{\beta_0^2}{\alpha_0^2\,(M_P\,c)^2}\frac{A\,m_p\,k_BT}{1-2\,\frac{\beta_0}{\alpha_0\,(M_P\,c)}\sqrt{2\,A\,m_p\,k_BT}}\right)}-1 \nonumber \\
&=&\!\!\ln{\left(\frac{\mathcal{G}_{1,2}^2}{1-\mathcal{G}_{1,2}}\right)}~, 
\end{eqnarray}
where
\begin{eqnarray}
\label{gvaril}
\mathcal{G}_{1,2}=1-\sqrt{1+4\,\frac{\beta_0}{\alpha_0^2}\left(\frac{\omega_{L_{1,2}}^{\mathrm{FWHM}}}{\omega_{B,0}}-1\right)}~.
\end{eqnarray}
The transcendental Eq. (\ref{transcendental}) can be numerically solved for $\mathcal{G}_{1,2}$. Note that one would have also arrived at Eq. (\ref{transcendental}), by using distribution Eq. (\ref{Cdistr}).
To solve Eq. (\ref{transcendental}), one needs to split it in two functions
\begin{eqnarray}
\!\!y_1\!\!&=&\!\!\frac{\alpha_0^2\,(M_P\,c)^2}{\beta_0^2}\,\frac{\mathcal{G}_{1,2}^2}{8\,A\,m_p\,k_BT} +\ln{\left(4\,\frac{\beta_0^2}{\alpha_0^2\,(M_P\,c)^2}\frac{A\,m_p\,k_BT}{1-2\,\frac{\beta_0}{\alpha_0\,(M_P\,c)}\sqrt{2\,A\,m_p\,k_BT}}\right)}-1 \\
\!\!y_2\!\!&=&\!\!\ln{\left(\frac{\mathcal{G}_{1,2}^2}{1-\mathcal{G}_{1,2}}\right)}~,
\end{eqnarray}
and plot them in the same graph. They are expected to intersect at two points, where $y_1=y_2$, which are the solutions for $\mathcal{G}_{1,2}$. An example of this is shown in Fig. \ref{csolut} for the relevant case of an ensemble of ${}^{129}\mathrm{Xe}$ atoms at $T=560\,\mathrm{K}$ and for $\alpha_0=10^8$, where $\beta_0=\alpha_0^2$ was assumed. 
It is then straightforward to obtain $\omega_{B_{1}}^{\mathrm{FWHM}}$ and $\omega_{B_{2}}^{\mathrm{FWHM}}$ from the obtained values of $\mathcal{G}_{1,2}$, using Eq. (\ref{gvaril}). The FWHM for the above ${}^{129}\mathrm{Xe}$ case turns out as $\mathrm{FWHM}_\omega\simeq1.5\times10^{-7}\,\mathrm{Hz}$. Note that the FWHM values for distributions from Eqs. (\ref{omdist}) and (\ref{Cdistr}) are related by $\mathrm{FWHM}_\omega=\omega_{B,0}\,\mathrm{FWHM}_C$. Therefore, one can also evaluate $\mathrm{FWHM}_C\simeq2\times10^{-15}$, which does not rely on the actual value of $\omega_{B,0}$.

\vspace{0.0cm}
\begin{figure}[h]
\hbox{\hspace{-0.5cm}\includegraphics[scale=0.207,trim={7.0cm 7.0cm 10.6cm 11.5cm},clip]{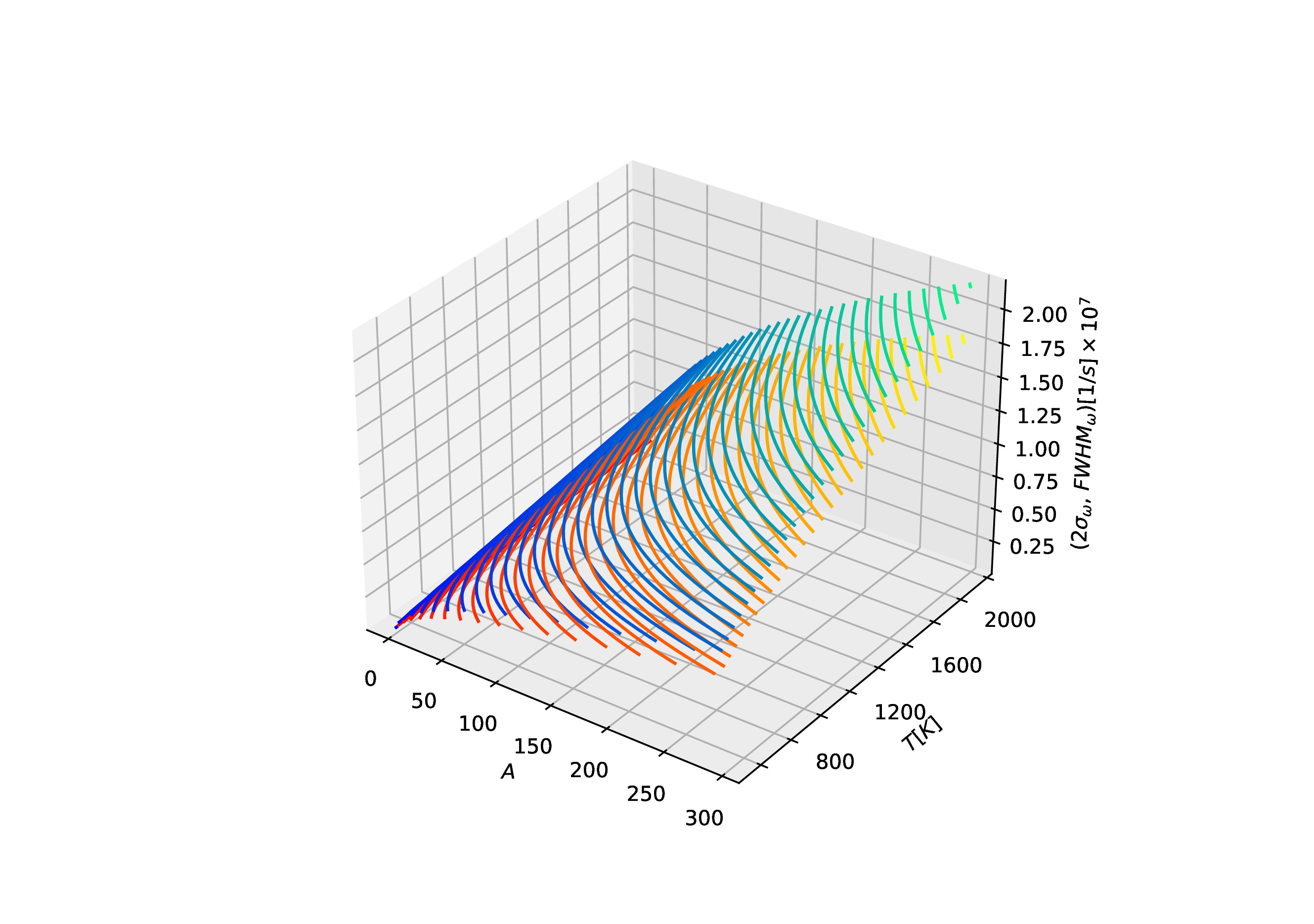}}
    \caption[Full width half maximum $\mathrm{FWHM}_\omega$ and standard deviation $\sigma_\omega$ of the Larmor frequency distribution]{Full width half maximum $\mathrm{FWHM}_\omega$ (blue-cyan) and standard deviation $\sigma_\omega$ (red-yellow) of the Larmor frequency distribution, as a function of atomic mass number $A$ and temperature $T$, at $\alpha_0=10^8$.}
    \label{sdf}
\end{figure}

The $\mathrm{FWHM}_\omega$ is obtained, as described above, for a range of atomic mass numbers $A$ and temperatures $T$, at the relevant $\alpha_0=10^8$, and shown in Fig. \ref{sdf}, alongside $2\,\sigma_\omega$ (see Eq. (\ref{smlin})). Note the factor $2$ in front of $\sigma_\omega$. It is there to provide a measure of width, since $\sigma_\omega$ alone is a measure of deviation from the average. One can see that both, $2\,\sigma_\omega$ and $\mathrm{FWHM}_\omega$ 
scale approximately as a square root in both, $A$ and $T$, and that they are of the same order of magnitude at a given parameter $\alpha_0$, where the $\mathrm{FWHM}_\omega$ is slightly greater. This is expected, since $\mathrm{FWHM}_\omega$ is the width of the distribution at half of its maximum value, and $2\,\sigma_\omega$ is the width of the distribution at a slightly higher value than that of half maximum, which makes it slightly narrower. For the projected experimental precision of $10^{-15}$, the line broadening for $\alpha_0=10^8$ is within reach of detection. 


\section{Non-Thermal Distribution of Atom Velocities}
\label{sec:nonthermalvel}

\begin{figure}[H]
    \hbox{\hspace{1.5cm}\includegraphics[scale=0.86,trim={3.6cm 3.6cm 1.6cm 4.8cm},clip]{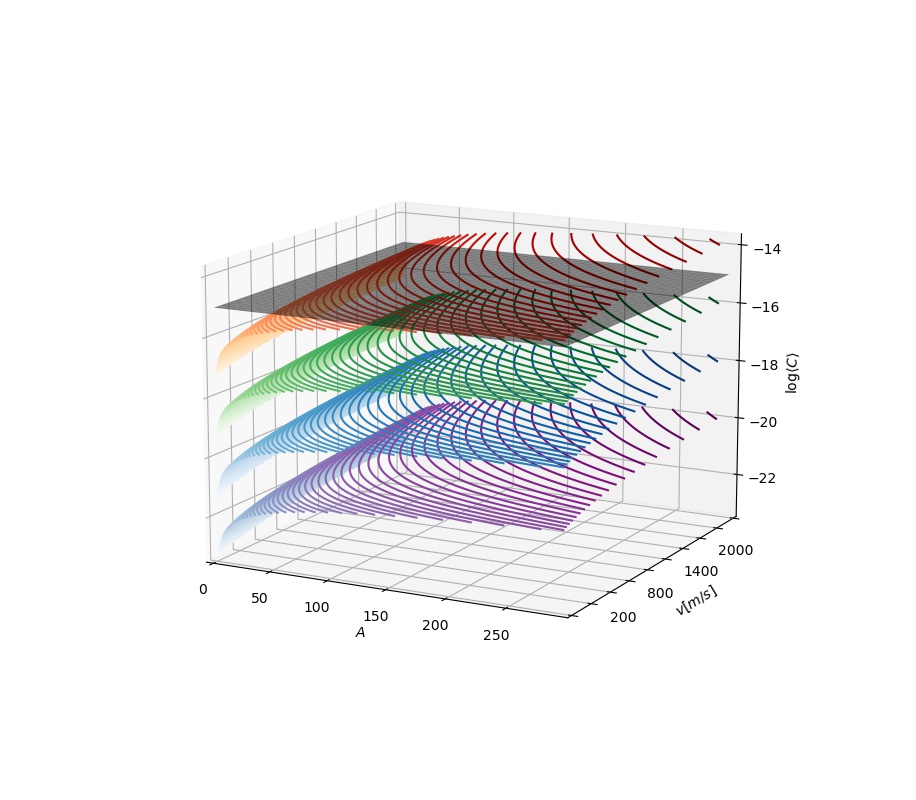}}
    \caption[Logarithmic dependence of $\langle C\rangle$ on mass number $A$ and atom velocity $v$ for four different values of $\alpha_0$]{Logarithmic dependence of $\langle C\rangle$ on mass number $A$ and atom velocity $v$ for four different values of $\alpha_0$; purple: $\alpha_0=10^2$, blue: $\alpha_0=10^4$, green: $\alpha_0=10^6$ and red: $\alpha_0=10^8$. The black flat surface is the experimental precision limit.}
  \label{logC12}
\end{figure}

In the previous sections, QG signatures induced by thermal distributions of ensembles of atoms were considered. While such considerations provide promising results, they are limited by the temperature dependence of QG signatures.
While these signatures increase with temperature, this is accompanied 
by increased thermal noise and the corresponding 
decrease of precision of the experiment. 
However, since QG signatures in Larmor frequencies depend on the velocities of atoms, one can introduce methods other than thermal motion, to control the velocities. For a non-thermal distribution of velocities, one can write the QG correction as
\begin{eqnarray}
\label{CAv}
\langle C\rangle(\alpha_0,\beta_0;\,A,v) =\alpha_0\,\frac{A\,m_p\,v}{M_P\,c}-\beta_0\,\frac{A^2m_p^2v^2}{(M_P\,c)^2}~.
\end{eqnarray}
For clarity, the logarithm of Eq. (\ref{CAv}) is shown in Fig. \ref{logC12}, as a monotonically increasing function of both $A$ and $v$, where $\beta_0=\alpha_0^2$ is assumed. The dependence is shown for different values of $\alpha_0$. The black flat surface corresponds to the experimental precision. One can again see that QG signatures can be detected if $\alpha_0\approx10^8$. In this case, the detection of QG signatures, using ${}^{129}\mathrm{Xe}$ atoms with $A=A_{Xe}=129$, would take place at $v\approx300\,\mathrm{m/s}$, given Eq. (\ref{CAv}). Note that Fig. \ref{logC12} is similar to Fig. \ref{logC11}. This is expected, since the same form of $\langle C\rangle$ is used, with the difference that in Fig \ref{logC11} the velocity is taken to be thermal.

One of the methods used to induce non-thermal motion of atoms is convection, where a current of atoms is passed through a duct. Once the flow is stationary, one can use the two-photon laser spectroscopy method, to measure the Larmor frequency of atoms. For the purposes of the proposed experiment, a velocity distribution of an incompressible, viscous fluid in a square duct is assumed \cite{dcts}
\begin{eqnarray}
\label{dvdist}
v=v_{max}\left[1-\left(\frac{x}{a\slash2}\right)^{2}\right]\left[1-\left(\frac{y}{a\slash2}\right)^{2}\right]~,
\end{eqnarray}
where $v_{{max}}$ is the maximum velocity in the centre of the duct and $a$ is the length of the inner side of the duct. Eq. (\ref{dvdist}) is then inserted in Eq. (\ref{larmorCorr}) to obtain the QG corrected Larmor frequency, as a function of $v_{max}$, and the position $(x,y)$ 
inside the duct as
\begin{eqnarray}
\label{veldistduct}
\omega_B(x,y)\!\!&=&\!\!\omega_{B,0}\left(1-\alpha_0\,\frac{A\,m_p\,v_{max}}{M_P\,c}\left[1-\left(\frac{x}{a\slash2}\right)^{2}\right]\left[1-\left(\frac{y}{a\slash2}\right)^{2}\right]\right. \nonumber \\
&{}&\left.\,\,\,\,\,\,\,\,\,\,\,\,\,\,\,\,\,\,\,\,\,\,+\beta_0\,\frac{A^2m_p^2\,v_{{max}}^2}{(M_P\,c)^2}\left[1-\left(\frac{x}{a\slash2}\right)^{2}\right]^2\left[1-\left(\frac{y}{a\slash2}\right)^{2}\right]^2\right)~.
\end{eqnarray}
A maximum QG induced deviation of the Larmor frequency is located at the centre of the duct, where $x=y=0$. In this case, $\langle C\rangle$ follows Eq. (\ref{CAv}) for $v=v_{max}$ (see Fig. \ref{logC12}). However, such measurements at exact locations are not possible in actual experiments. Therefore, two possible scenarios are considered in the following, namely an average measurement in a line perpendicular to the duct and an average measurement in the total cross section of the duct.

In the case where the Larmor frequency is measured in a line which runs perpendicularly through the duct, one needs to average Eq. (\ref{veldistduct}) over either of the $(x,y)$ dimensions as
\begin{eqnarray}
\label{omx}
\omega_B(x)\!\!&=&\!\!\frac{1}{a}\int_{-a/2}^{a/2}\omega_B(x,y)\,\mathrm{d}y \nonumber \\
&=&\!\!\omega_{B,0}\left(1-\alpha_0\,\frac{2\,A\,m_p\,v_{max}}{3\,M_P\,c}\left[1-\left(\frac{x}{a\slash2}\right)^{2}\right]\right. \nonumber \\
&{}&\,\,\,\,\,\,\,\,\,\,\,\,\,\,\,\,\,\,\,\,\,\,\left.+\beta_0\,\frac{8\,A^2m_p^2\,v_{max}^2}{15\,(M_P\,c)^2}\left[1-\left(\frac{x}{a\slash2}\right)^{2}\right]^2\right)~,
\end{eqnarray}
from where one can see that the maximum deviation is obtained for $x=0$. In the case where the Larmor frequency is measured over the whole cross section area of the duct, then one needs to average Eq. (\ref{veldistduct}) over both, $x$ and $y$, as
\begin{eqnarray}
\label{omxy}
\bar{\omega}_L\!\!&=&\!\!\frac{1}{a^2}\int_{-a/2}^{a/2}\int_{-a/2}^{a/2}\omega_B(x,y)\,\mathrm{d}x\,\mathrm{d}y \nonumber \\
&=&\!\!\omega_{B,0}\left(1-\alpha_0\,\frac{4\,A\,m_p\,v_{max}}{9\,M_P\,c}+\beta_0\,\frac{64\,A^2m_p^2\,v_{max}^2}{225\,(M_P\,c)^2}\right)~.
\end{eqnarray}
One can see from Eqs. (\ref{omx}) and (\ref{omxy}) that in either choice of measurements, QG signatures will approximately be described by Eq. (\ref{CAv}). The only difference is in the value of the particle velocity, which is of order $v\approx v_{max}$. At this time it is difficult to determine which of the above two applications would work best in an actual experimental setup, due the non-linear dependence of the atomic transition rate on intensity, and the dependence on the number of atoms in the illuminated volume \cite{magnetometer2023}.

To discuss the effects on the experimental sensitivity, a larger velocity can be sampled (see the last paragraph of Section IV. A. in Ref. \cite{magnetometer2023}). For example, scaling the velocity by a factor of 2, $\alpha_{0,min}$ 
is then scaled by  $2^{3/4}\sim1.7$. The beam waists considered in Ref. \cite{magnetometer2023} are $w_0 = 0.1 - 1\,\mathrm{mm}$ and the transit-time-limited FWHM is given by $\delta\nu =  0.4\,v / w$ \cite{Demtroder2002}.
If the velocity is $v = 400\,\mathrm{m/s}$, (corresponding to a temperature $T = 1200\,\mathrm{K}$ in the thermally induced velocities case), and the beam waist is scaled by a factor of $1/2^{1/4} \sim 0.84$, the transit time broadening falls in the $2 - 0.2\,\mathrm{MHz}$ range, which is negligible in the Doppler broadened case ($1.26\,\mathrm{GHz}$).
If the beam waist is not scaled as such, the transit time broadening is negligibly reduced.

\section{Conclusion}
\label{sec:conc}

Magnetometers are highly advanced experimental apparatuses, which can be designed to measure Larmor frequencies of atoms in an external magnetic field with unprecedented precision. This makes a magnetometer experiment an ideal candidate to search for QG signatures. Such signatures are predicted through GUP motivated QG modifications of the Larmor frequency of an atom in an external magnetic field \cite{Bosso:2016frs}. A specific experimental proposal is considered, where the Larmor frequency of ${}^{129}\mathrm{Xe}$ atoms in an external magnetic field $B=2.7\,\mathrm{\mu T}$ is measured.

Given the formulation of QG modifications of the Larmor frequency, it is proposed that such modifications manifest through relative velocities of individual atoms. A natural way to induce such velocities is to consider a thermalized ensemble of atoms. This means they follow a certain velocity distribution at a given temperature. It turns out that the Maxwell-Boltzmann distribution is the relevant velocity distribution for the proposed experiment. This gives rise to a QG induced distribution of Larmor frequencies of an ensemble of atoms, given by Eq. (\ref{omdist}). Such a distribution causes a deviation of the average Larmor frequency from its standard prediction, given by Eq. (\ref{omcorr}) (see also Fig. \ref{logC11}). Furthermore, it causes a distribution of deviations from the standard prediction, which implies a width of the QG signature, described by the standard deviation or the FWHM, given by Eqs. (\ref{smlin}) and (\ref{fwhmgupmag}), respectively (see also Figs. \ref{cdist1} and \ref{sdf}).

On the other hand, velocities of atoms can also be induced by non-thermal methods. One of such methods is using convection currents. In this case, a gas of atoms is passed through a duct at a controlled velocity. For the proposed experiment, a square duct is considered and a velocity distribution of an incompressible, viscous fluid is assumed. Since the experiment can be designed to measure the Larmor frequency either in a line through the cross section of the duct, or the whole cross section of the duct, one can predict the corresponding QG signatures, given by Eqs. (\ref{omx}) and (\ref{omxy}), respectively (see also Fig. \ref{logC12}).

The projected precision of the proposed magnetometer experiment suggests that QG signatures in Larmor frequency measurements can be observable for $\alpha_0\approx10^{8}$ (see Figs. \ref{logC11} and \ref{cdist1} and \ref{logC12}). In the above, one assumes $\beta_0=\alpha_0^2$. Since this estimate is well below $\alpha_{EW}$, there is a strong chance to detect QG signatures for the first time. In case no such signatures are observed, an unprecedented bound of $\alpha_0<10^8$ will be set.

The predicted QG signatures by either thermal or non-thermal velocities are promising. It turns out that the Larmor frequency of an atom obtains QG corrections at non-vanishing velocities. In this section, detailed considerations were provided in the case of thermal movement of atoms, described by the MB distribution, and in the case of convection currents. Given the projected experimental precision of the magnetometer, using ${}^{129}\mathrm{Xe}$ atoms, one can observe QG signatures for $\alpha_0=10^8$ in both cases. However, if QG signatures are not observed, this experiment will set a strong bound $\alpha_0<10^8$. It will improve the electroweak bound, set by $\alpha_{EW}=10^{17}$, by nine orders of magnitude!

\section*{Competing interests}

The authors declare no competing interests.

\section*{Data availability}

Data sharing is not applicable to this article as no new data were created or analyzed in this study.

\end{document}